\documentclass[]{spie}
\usepackage[]{graphicx}
\usepackage{amsmath}
\usepackage{amsfonts}
\usepackage{bm}
\usepackage{color}

\title{
Coherence and Entanglement in Two-Qubit Dynamics: Interplay of the
Induced Exchange Interaction and Quantum Noise due to Thermal
Bosonic Environment
}

\author{\textbf{Vladimir Privman} and \textbf{Dmitry Solenov}
\skiplinehalf %
Department of Physics, Clarkson
University, Potsdam, NY 13699--5721, USA}

\authorinfo{Electronic addresses: Privman@clarkson.edu, Solenov@clarkson.edu}

\newcommand{\ind}[1]{\textrm{#1}}
\newcommand{\com}[1]{}

\newcommand{\Fig}[2]
 {
    \begin{figure}
   \begin{center}
   \begin{tabular}{c}
   \includegraphics[height=5.5cm]{#1}
   \end{tabular}
   \end{center}
   \caption[example]
   { \label{#1} #2}
\vphantom{.}
\hrule
   \end{figure}

 }

\newcommand{\red}[1]{%
{#1}}

\begin{document}

\maketitle

\begin{abstract}

We present a review of our recent results for the comparative
evaluation of the induced exchange interaction and quantum noise
mediated by the bosonic environment in two-qubit systems. We
report new calculations for P-donor-electron spins in Si-Ge type
materials. Challenges and open problems are discussed.

{\noindent}\red{{\small{}{\bf Keywords:}\ decoherence, entanglement, qubit, exchange interaction, quantum noise, quantum computing.}}

\vphantom{.}
\hrule
\end{abstract}
\

\section{INTRODUCTION}\label{sec:intro}

In this review we present results of our recent
investigations\cite{STPs,STPb} addressing several problems in the
field of open quantum systems; some have a rather long history.
With the actual experimental probes now being carried out at the
nanoscale, these problems have become more pressing, and some have
actually been suggested by the experimental
developments.\cite{Jiang,Jiang2,Craig,Elzerman,Koppens,Petta,MMJ,PF,Nakamura,Vion,Chiorescu,Yamamoto}
Perhaps the most fundamental (and difficult) of these problems is
the matter of accounting within a calculationally tractable
approach for relaxation vs.\ coherent dynamics in open quantum
systems. In particular the coherent and quantum-noise features
induced by the bosonic environment (bath of noninteracting bosonic
modes) are especially interesting, because this model is widely
applicable for quantum computing systems. Here we consider
bath-induced RKKY-type exchange interactions in two-qubit systems.

This paper is organized as follows. In the next section, Sec.~\ref{sec:M}, we
formulate the model and discuss the issue of the initial
conditions. An example of the semiconductor (Si-Ge type) based qubit
model is given and the nature of the interactions is discussed.

Section~\ref{sec:onset} is devoted to the initial dynamics of the
induced interaction and quantum noise. In particular, in
Sec.~\ref{ssec:OS} we present an exactly solvable model to obtain
the exchange interaction as well as the time-dependent correction
due to the initial conditions. The model also demonstrates the delay in
the response of the qubits caused by the finite propagation
velocities of the mediating \red{virtual} bosons. In Sec.~\ref{ssec:DM}, we study
the interplay of the coherent dynamics and quantum noise by calculating the
concurrence.\cite{Wootters1,Wootters2} Evolution of the reduced
density matrix elements is also investigated.

At large times, as the system forgets the initial state, the
exchange interaction becomes stationary. This regime is discussed
in Sec.~\ref{sec:LT}. We utilize the master-equation formulation
of Markovian dynamics, Sec.~\ref{ssec:ME}, and go over an
illustrative one-dimensional (1D) example of the emerging
interaction and quantum noise in Sec.~\ref{ssec:1D}.
Generalization to higher dimensionality is given in
Sec.~\ref{ssec:GD}. A solid-state based three-dimensional (3D)
example of a semiconductor qubit model is presented in
Sec.~\ref{ssec:3D}. In the latter discussion we compare
bath-induced coupling and noise to other interactions, such as an
electromagnetic coupling of P-impurity spins (qubits).
Finally, brief summary and outline of some open
challenges are offered in Sec.~\ref{sec:discuss}.

\hphantom{A}

\section{THE MODEL AND INITIAL CONDITIONS}\label{sec:M}

We consider two qubits, i.e., two two-state quantum systems,
modeled by localized electron spins ($1/2$) immersed in a
common quantum environment. The qubits are located at the distance
$\mathbf{d}$ from each other, far enough so that the direct
overlap of \red{the electron wave functions} is negligible. The environment is
modeled by a bath of bosonic modes, which are maintained at
temperature $T$. It is usually assumed that external
influences, as well as possibly
internal bath-mode interactions, set a fast time scale $t_B$ for the bath correlations to
equilibrate. The bath modes are then regarded as otherwise noninteracting. At least for the low-frequency bath modes, it is
usually argued that such a thermalization time for a generic case should be of
order $\hbar/k_{\textrm{B}}T$. The thermal-state density matrix of the bath modes, taken
here as noninteracting bosonic fields, is
\begin{equation}\label{Eq:M:rhoB}
\rho_B=\mathrm{Z}^{-1} \exp \bigg[{-{\sum\limits_{\mathbf{k},\mathbf{\xi}}
{\omega_{\mathbf{k},\mathbf{\xi}} a_{\mathbf{k},\mathbf{\xi}}^\dag
a_{\mathbf{k},\mathbf{\xi}}} }/{k_{\textrm{B}}T}}\bigg] \, ,
\end{equation}
where we set $\hbar = 1$, and the partition function is $\mathrm{Z}=1/\prod_k
(1-e^{-\omega_k/k_{\textrm{B}} T})$. The bath is linearly coupled to each
qubit,
\begin{equation}\label{Eq:M:H_SB}
H_{SB}=\sum\limits_{j=1,2}\sum\limits_{m=x,y,z}\sigma^j_m X^j_m \, ,
\end{equation}
where the superscripts ($j=1,2$) label the two spins, and the bath operators are given by
\begin{equation}\label{Eq:M:X_jm}
X^j_m=\sum_{\mathbf{k},\mathbf{\xi}}g^m_{\mathbf{k},\mathbf{\xi}}e^{i\mathbf{k}\cdot\mathbf{r}_j}
\left(a_{\mathbf{k},\mathbf{\xi}}+a_{-\mathbf{k},\mathbf{\xi}}^{\dagger}\right)\, .
\end{equation}

The overall Hamiltonian of the qubit-bath system is
\begin{equation}\label{Eq:M:H}
H = H_S + H_B + H_{SB}\, ,
\end{equation}
where
\begin{equation}\label{Eq:M:H_B}
H_B=\sum_{\mathbf{k},\mathbf{\xi}}\omega
_{\mathbf{k},\mathbf{\xi}}a_{\mathbf{
k},\mathbf{\xi}}^{\dagger}a_{\mathbf{k},\mathbf{\xi}} \, ,
\end{equation}
and the two spin-$1/2$ (qubits) will be assumed split by external
magnetic field,
\begin{equation}\label{Eq:M:H_S}
H_S=\Delta (\sigma_z^1 + \sigma_z^2 )/2 \, .
\end{equation}
Here $\Delta$ is the energy gap between the up and down states for
spins 1 and 2. A natural example of such a system are spins
of two localized electrons interacting via lattice vibrations
(phonons) by means of the spin-orbit
interaction.\cite{Mahan,Hasegawa,Roth,SO-Winkler} Another example
is provided by atoms or ions in a cavity, used as two-state
systems interacting with photons.\cite{PGCZ}

Our emphasis here will be on calculating and comparing the
relative importance of the coherent (induced interaction) vs.\
quantum-noise effects of a given bosonic bath in the two-qubit
dynamics. We do not include other possible two-qubit interactions
in such comparative calculation of dynamical quantities. \red{In
Sec.~\ref{ssec:3D} we also include the direct electromagnetic (EM)
coupling for a comparison of the bath-induced and EM dipole-dipole
interaction strengths.}

An interesting realization of the model formulated above can be
found in the impurity electron spin dynamics in semiconductors.
Let us consider P donor impurities embedded at controlled
positions in an otherwise very clean Si (or Ge) crystal
matrix. The system is maintained at very low temperature,
as appropriate for quantum computing. Therefore, the outer
donor-impurity electron remains bound. The spins of such localized
electrons can be utilized as
qubits. They are subject to external magnetic field
$\mathbf{H}$,
which produces the Zeeman splitting (\ref{Eq:M:H_S}). The
spin-orbit interaction couples the spin to the deformation
potential fluctuations of the host semiconductor, producing the
energy change
\begin{equation}\label{Eq:M:H-SO}
H_\textrm{SO}=\mu_{\textrm{B}} \! \! \sum\limits_{m,l\,=\,x,y,z}
\! \! {\sigma_m g_{ml}H_l}\,,
\end{equation}
where $\mu_\textrm{B}$ is the Bohr magneton, and
$\mathbf{H}=\{H_x,H_y,H_z\}$. Here the tensor $g_{ml}$ is
sensitive to lattice deformations. It was shown\cite{Roth} that
for the donor state which has tetrahedral symmetry (which is the
case for P in Si or Ge), the Hamiltonian (\ref{Eq:M:H-SO}) yields
the spin-deformation interaction of the form
\begin{equation}
H_i = {\rm A}\mu _{\textrm{B}} \left[ { \bar \varepsilon _{xx}
\sigma _x H_x + \bar \varepsilon _{yy} \sigma _y H_y  +  \bar
\varepsilon _{zz} \sigma _z H_z  + {\bar\Delta }
(\mathbf{\sigma}\cdot\mathbf{H}})/3\right]
\\ \label{Eq:M:H-SD-Hxyz}
+\,{\rm B}\mu _{\textrm{B}} \left[ {\bar \varepsilon _{xy} \left({
\sigma_x H_y+\sigma_y
H_x}\right)+{\text{c}}{\text{.p}}{\text{.}}}\right]\,.
\end{equation}
Here c.p.\ denotes cyclic permutations and $\bar\Delta$ is the
effective dilatation. The tensor $\bar\varepsilon_{ij}$ already
includes averaging of the strain with the gradient of the
potential over the donor ground state wave function.

As before, let us assume that we have two impurities separated by
distance \textbf{d}. For definiteness, one can direct the magnetic
field along the $z$-axis. Then the spin-deformation interaction
Hamiltonian simplifies to
\begin{equation}\label{Eq:M:H-SD-Hz}
H_i = {\rm A}\mu _{\textrm{B}}  \bar \varepsilon _{zz} \sigma _z^i
H_z + {\rm B}\mu _{\textrm{B}} \left( {\bar \varepsilon _{yz}
\sigma _y^i H_z  + \bar \varepsilon _{zx} \sigma _x^i H_z }
\right)\,.
\end{equation}
In terms of the quantized phonon field, we have\cite{Mahan,MKGB}
\begin{equation}\label{Eq:M:StrainTensor}
\bar \varepsilon _{ij}  = \sum\limits_{{\mathbf{k}},{\mathbf{\xi
}}} {f({\mathbf{k}})\sqrt {\frac{\hbar } {{8\rho V\omega
_{{\mathbf{k}},{\mathbf{\xi }}} }}} \left( {\xi _{{\mathbf{k}},i}
k_j  + \xi _{{\mathbf{k}},j} k_i } \right)\left(
{a_{{\mathbf{k}},{\mathbf{\xi }}}^\dag   +
a_{{\mathbf{k}},{\mathbf{\xi }}} } \right)}\,,
\end{equation}
where in the spherical donor ground state
approximation\cite{Hasegawa,MKGB}
\begin{equation}\label{Eq:M:f(k)}
f({\mathbf{k}}) = \frac{1} {{\left( {1 + a_\textrm{B}^2 k^2 }
\right)^2}}\,.
\end{equation}
Here $a_\textrm{B}$ is {\it half\/} the effective Bohr radius of
the donor ground state wave function. In an actual Si or Ge
crystal, donor states are more complicated and include corrections
due to the symmetry of the crystal matrix including the fast
Bloch-function oscillations. However, the wave function of the
donor electrons in our case is spread over several atomic
dimensions (see below). Therefore, it suffices to consider
``envelope'' quantities. Thus, in the spin-phonon Hamiltonian
(\ref{Eq:M:H_SB})-(\ref{Eq:M:X_jm}) coupling constants will be
taken in the form
\begin{equation}\label{Eq:M:gk-SD}
g_{{\mathbf{k}},{\mathbf{\xi }}}^m  = \frac{D_m}{{\left( {1 +
a_\textrm{B}^2 k^2 } \right)^2}}\sqrt {\frac{\hbar } {{8\rho
V\omega _{{\mathbf{k}},{\mathbf{\xi }}} }}} \left( {\xi
_{{\mathbf{k}},z} k_m  + \xi _{{\mathbf{k}},m} k_z } \right)\,,
\end{equation}
where $D_x=D_y={\rm B}\mu_\textrm{B} H_z$ and $D_z={\rm
A}\mu_\textrm{B} H_z$.

It will be instructive to consider a one-dimensional calculation,
which simplifies the notation. \red{General results\cite{STPs,STPb} are presented later
in this article.} One can think of a 1D channel geometry along the
$z$ direction. This will give an example of an Ohmic bath model
discussed later. In a 1D channel the boundaries\cite{PBS} can
approximately quantize the spectrum of the phonons along $x$ and
$y$, depleting the density of states except at certain resonant
values. Therefore, the low-frequency effects, including the
induced coupling and quantum noise, will become effectively
one-dimensional, especially if the effective gap due to the
confinement is of the order of $\omega_c$. The frequency cutoff
comes from (\ref{Eq:M:f(k)}), namely, it is due to the bound
electron wave function localization. A channel of width comparable
to $\sim a_\textrm{B}$ will be required. This, however, may be
difficult to achieve in Si or Ge with the present-day technology.
Other systems may offer more immediately available 1D geometries
for testing similar theories, for instance, carbon nanotubes,
chains of ionized atoms suspended in ion
traps,\cite{Leibfried,Marquet,Porras} etc. In the 1D case, the
longitudinal acoustic (LA, $||$) phonons will account for the $g_{\mathbf{k},\parallel}^z$
component of the coupling, whereas the transverse acoustic (TA,
$\perp$) phonons will affect only the $x$ and $y$ spin
projections.

It will be shown later that the contributions of the
cross-products of coupling constants,
$g_{\mathbf{k},\mathbf{\xi}}^m(g_{\mathbf{k},\mathbf{\xi}}^{m'})^*$
with $m\neq m'$, to quantities of interest vanish. The
combinations that enter the dynamics are
\begin{equation}\label{Eq:M:gk-1D}
|g_{k_z }^z|^2 = \frac{\textrm{A}^2 \mu_\textrm{B}^2 H_z^2}
{{4\rho V\omega _{k_z ,\parallel } }}\frac{{k_z^2 }} {{\left( {1 +
a_\textrm{B}^2 k_z^2 } \right)^4 }}\,,
\qquad
|g_{k_z }^x |^2  = |g_{k_z }^y |^2  = \frac{\textrm{B}^2
\mu_\textrm{B}^2 H_z^2} {{4\rho V\omega _{k_z , \bot }
}}\frac{{k_z^2 }} {{\left( {1 + a_{\rm B}^2 k_z^2 } \right)^4 }}\,.
\end{equation}
With the usual assumption for the low-frequency dispersion
relations $\omega _{k_z ,\parallel}\approx c_\parallel k_z$ and
$\omega _{k_z ,\perp}\approx c_\perp k_z$, the expressions
(\ref{Eq:M:gk-1D}) lead to the Ohmic bath model. The shape of the
frequency cutoff resulting from (\ref{Eq:M:gk-1D}) is not
exponential. However, to estimate the magnitude of the interaction
one can equivalently consider the exponential
cutoff,\cite{Leggett} with the summation over bosonic modes
carried out as
\begin{equation}\label{Eq:M:DOS}
\sum_{\mathbf{k},\xi}|g^m_{\mathbf{k},\xi}|^2 \to
\int\limits_0^\infty d\omega |g^m (\omega )|^2\Upsilon (\omega )=
\int\limits_0^\infty d\omega \ \alpha^m_n\omega^n
\exp(-\omega/\omega_c)\,,
\end{equation}
where $\Upsilon(\omega )$ is the density of states. The coupling
constants should then be taken as
\begin{equation}\label{Eq:M:alfas-1D}
\alpha_1^z = \frac{\textrm{A}^2 \mu_\textrm{B}^2 H_z^2}{8\pi
\rho S c^3_\parallel}\,,
\qquad\qquad
\alpha_1^x  = \alpha_1^y  = \frac{\textrm{B}^2 \mu_\textrm{B}^2
H_z^2}{8\pi \rho S c^3_\bot}\,,
\end{equation}
where $S$ is the cross section of the channel, and the cutoff is
$\omega_c \rightarrow c_{\parallel}/a_\textrm{B}$ for the $z$
component, and $\omega_c \rightarrow c_{\perp}/a_\textrm{B}$ for
the $x$ and $y$ components.

For numerical estimates, we note that a typical
value\cite{Hasegawa,Roth,MKGB} of the effective Bohr radius in Si
for the P-donor-electron ground state wave function is
$2a_\textrm{B}=2.0\,$nm. The crystal lattice density is
$\rho=2.3\times10^3\,$kg/m$^3$, and the g-factor is $g^*=1.98$. For
an order-of-magnitude estimate, we take a typical value of the
phonon group velocity, $c_s=0.93 \times 10^4\,$m/s. The spin-orbit
coupling constants in Si are\cite{Hasegawa,Roth,MKGB}
$\textrm{A}^2\approx 10^2$ and $\textrm{B}^2\approx10^{-1}$. In
the Ge lattice, the spin-orbit coupling is dominated by the
non-diagonal terms,\cite{Hasegawa,Roth,MKGB} $\textrm{A}^2\approx
0$ and $\textrm{B}^2\approx10^6$. The other parameters are
$2a_\textrm{B}=4.0\,$nm, $\rho=5.3\times10^3\,$kg/m$^3$,
$c_s=5.37\times10^3\,$m/s, and $g^* = 1.56$. This results in a
much stronger transverse component interaction. In both cases the
magnetic field will be taken of order $H_z=3\times 10^4\,$G. One could
use other experimentally suggested
values for the parameters, such as, for instance, $a_{\rm B}$. This will
not affect the results significantly.

As mentioned earlier, a realistic phonon environment includes the
time scale at which it is reset to the thermal state. One can
think of the short- and long-time dynamics in reference to this
time scale. In the long-time regime, it is customary to introduce
Markovian-type assumptions, which include resetting the bath to
the thermal state instantaneously. We will use this approach in
the master-equation formulation of the large time dynamics,
Sec.~\ref{ssec:ME}.

For the short-time dynamics, one has to address the matter of the
initial conditions at $t=0$. We will use the initially factorized
density matrix, with the thermal state assumed for the bath. In
quantum computing applications, such an initial condition has been
widely used for the qubit-bath
system,\cite{Privman,Tolkunov2,Solenov}
\begin{equation}\label{Eq:M:IC}
\rho(0) = \rho_S(0)\rho_B\,.
\end{equation}
This choice allows comparison with the Markovian results, and is
usually needed in order to make the short-time approximation
schemes tractable;\cite{Solenov} specifically, it is necessary for
the exact solvability of the model considered in the next section.

A somewhat more ``physical'' excuse for the factorized initial
conditions is formulated as follows. Quantum computation is carried
out over a sequence of time intervals during which various
operations are performed on individual qubits and on pairs of
qubits. These operations include control gates, and measurements for
error correction. It is usually assumed that these ``control''
functions, involving rather strong interactions with external
objects, as compared to interactions with sources of quantum
noise, erase the fragile entanglement with the bath modes that
qubits can develop before those time intervals when they are
``left alone'' to evolve under their internal (and bath induced)
interactions. Thus, for evaluating relative importance of the
quantum noise effects on the internal (and bath induced) qubit
dynamics, which is our goal here, we can assume that the state of the
qubit-bath system is ``reset'' to uncorrelated at $t=0$.

\hphantom{A}

\section{ONSET OF CORRELATIONS BETWEEN QUBITS}\label{sec:onset}

Let us consider relatively short time scales and analyze how the
exchange interaction, accompanied by the quantum noise, sets in.
One can argue that such an investigation is only feasible provided
one knows the initial condition for the qubits-bath density matrix.
Indeed, the short-time interaction (and quantum noise) should depend on the
initial condition. The factorized initial condition, with the thermal-state bath, assumed
here, is expected in most quantum computing applications, as
discussed above.

\subsection{Development of Induced Exchange Interaction}\label{ssec:OS}

When the evolution of isolated qubits is slow with respect to the
other time scales such as that of decoherence, so that one can
assume vanishing qubit splitting energy, $\Delta =0$, and
if only one system operator enters the
qubit-bath interaction, then one obtains an exactly solvable model. A more general ``adiabatic'' system that allows exact solvability is obtained when $H_{S}$ commutes with $H_{SB}$. For
example, for electron spins of P-impurities in Si one has the spin-phonon
interaction dominated by such an adiabatic term, and only one such system operator in the
interaction matters for the onset dynamics. Though this situation
is rarely the case in quantum computing systems, the present model provides
a convenient tool for evaluating the initial dynamics of the exchange
interaction build-up, as well as for analyzing the response delay
due to the finite speed of the mediating \red{virtual} bosonic particles. These
features are difficult to capture analytically in other models.

Therefore, let us presently take $\alpha _n^y=\alpha _n^z=0$,
while $\alpha _n^x \neq 0$. With the above assumptions, one can
utilize the bosonic operator techniques\cite{Louisell} to obtain
the reduced density matrix for the system (\ref{Eq:M:H_SB})-(\ref{Eq:M:X_jm}),
\begin{equation}\label{Eq:OS:AdiabaticSolution}
\rho _S (t) = \sum\limits_{\lambda ,\lambda '} {P_\lambda  \rho _S
(0)P_{\lambda '} e^{\frak{L}_{\lambda \lambda '} (t)} }\,,
\end{equation}
where the projection operator is defined as
$P_\lambda=\left|{\lambda_1\lambda_2}\right\rangle\left\langle
{\lambda _1\lambda _2}\right|$, and
$\left|{\lambda_j}\right\rangle$ are the eigenvectors of
$\sigma_x^j$. The exponent in (\ref{Eq:OS:AdiabaticSolution})
consists of the real part, which leads to decay of off-diagonal
density-matrix elements resulting in decoherence,
\begin{equation}
  \textrm{Re} \, \frak{L}_{\lambda \lambda '}(t)= -\sum\limits_k {G_k (t,T)
  \bigg[ {\left( {\lambda '_1  - \lambda _1 } \right)^2  + \left({
  \lambda '_2  - \lambda _2 } \right)^2 } }
\label{Eq:OS:DecoherenceFunction-ReL}
   + \,{2\cos\! \left( {\frac{{\omega _k \left| {\mathbf{d}} \right|}}
{{c_s }}} \right)\!\left( {\lambda '_1  - \lambda _1 }
\right)\left( {\lambda '_2  - \lambda _2 } \right)} \bigg]\, ,
\end{equation}
and the imaginary part, which describes the coherent evolution,
\begin{equation}\label{Eq:OS:DecoherenceFunction-ImL}
\textrm{Im} \,\frak{L}_{\lambda \lambda '} (t) =  \sum\limits_k {C_k
(t)\cos \left( {\frac{{\omega _k \left| {\mathbf{d}} \right|}}
{{c_s }}} \right)\left( {\lambda _1 \lambda _2  - \lambda '_1
\lambda '_2 } \right)}\, .
\end{equation}
Here we defined the standard spectral
functions\cite{Leggett,Privman}
\begin{equation}\label{Eq:OS:DecoherenceFunction-Gk}
G_k (t,T) = 2\frac{{\left| {g_k } \right|^2 }} {{\omega _k^2
}}\sin ^2 \left( \frac{{\omega _k t}} {{2}}\right) \coth \left(
{\omega _k \over 2k_{\textrm{B}} T}  \right)
\end{equation}
and
\begin{equation}\label{Eq:OS:DecoherenceFunction-Ck}
C_k (t) = 2\frac{{\left| {g_k } \right|^2 }} {{\omega _k^2
}}\left( {\omega _k t - \sin \omega _k t} \right)\,.
\end{equation}
Calculating the sums by converting them to integrals over the
bath-mode frequencies $\omega$ in
(\ref{Eq:OS:DecoherenceFunction-ReL}) and
(\ref{Eq:OS:DecoherenceFunction-ImL}), assuming the Ohmic bath
with $n=1$, for $T>0$ one obtains a linear in time, $t$,
large-time behavior for both the temperature-dependent real part
and for the imaginary part. The coefficient for the former is
$\sim\! kT$, whereas for the latter it is $\sim\! \omega_c$. For
the super-Ohmic models, $n>1$, the real part grows slower, as was
also noted in the literature.\cite{Privman,PALMA,VKampen,Hanggi}

\Fig{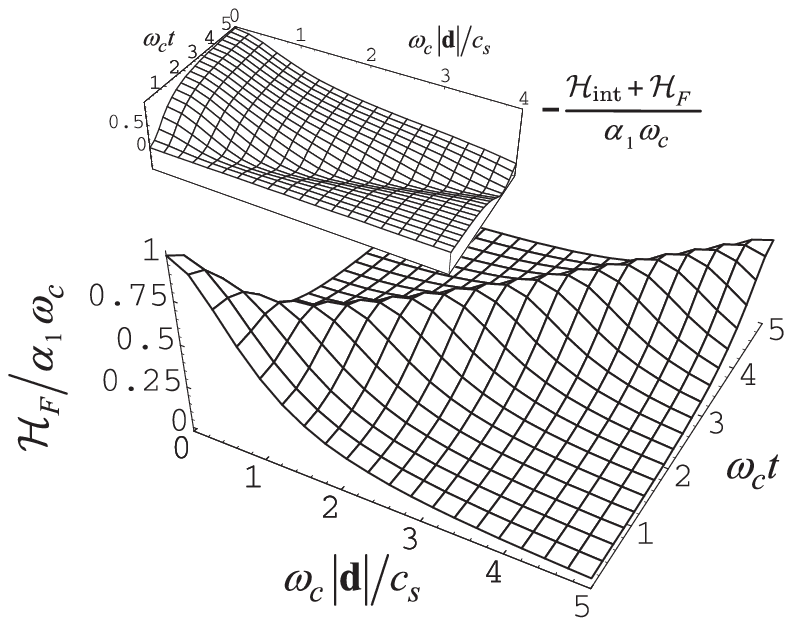}{The magnitude of the time-dependent Hamiltonian
corresponding to the initial correction as a function of time and
distance.  The Ohmic ($n=1$) case is shown. The inset demonstrates
the onset of the cross-qubit interaction on the same time scale.}

First, let us analyze the coherent evolution part in
(\ref{Eq:OS:AdiabaticSolution}), namely, the effect that the
imaginary part of $\frak{L}_{\lambda \lambda '}(t)$ has on the
evolution of the reduced density matrix, since this contribution
leads to the induced interaction. If we omitted
(\ref{Eq:OS:DecoherenceFunction-ReL}) from
(\ref{Eq:OS:AdiabaticSolution}),
(\ref{Eq:OS:DecoherenceFunction-ImL}), and
(\ref{Eq:OS:DecoherenceFunction-Ck}), we would
obtain the (coherent) evolution operator in the form
\begin{equation}
e^{-i\left(H_\textrm{int}+F(t)\right)t}\;.
\end{equation}
The interaction $H_\textrm{int}$ comes from the first term in
(\ref{Eq:OS:DecoherenceFunction-Ck}),
\begin{equation}\label{Eq:OS:H-int}
H_\textrm{int}\! = \! - \frac{2\alpha _n^x \Gamma (n)c_s^n\omega
_c^n } {\left(c_s^2 + \omega _c^2 \left| {\mathbf{d}} \right|^2
\right)^{n/2} } \cos \left[{
 n\arctan \left( {\frac{{\omega _c \left| {\mathbf{d}} \right|}}
{{c_s }}} \right)} \right]\!\sigma _x^1 \sigma _x^2\,.
\end{equation}
This expression gives the constant interaction that is important
in large time dynamics. We will obtain this induced exchange
interaction latter using different techniques for more general
cases, see Sec.~\ref{sec:LT}. The operator $F(t)$ is given by
\begin{equation}\label{Eq:OS:F(t)}
F(t) = 2\sigma _x^1 \sigma _x^2 \int\limits_0^\infty  {d\omega
\frac{{D(\omega )\left| {g(\omega )} \right|^2 }} {\omega
}\frac{{\sin \omega t}} {{\omega t}}\cos \left( \frac{\omega
|\mathbf{d}|}{c_s} \right)} \, .
\end{equation}
It commutes with $H_\textrm{int}$ (and with itself at different
times), and therefore $d\left(tF(t)\right)/dt$ can be viewed as
the initial time-dependent correction to the interaction. In fact,
this term controls the onset of the induced coherent interaction;
note that $F(0)=-H_\textrm{int}$, but for large times $F(t) \sim
\alpha^x_n \omega_c^n/(\omega_c t)^n$.

Let us consider in detail the time dependent correction
$H_F(t)=d\left(tF(t)\right)/dt$ to the interaction Hamiltonian during the initial
evolution,
\begin{equation}\label{Eq:OS:H_F}
H_F(t)= \sigma _x^1 \sigma _x^2 \alpha_n\Gamma(n)
\Big[
u(\omega_c|\mathbf{d}|/c_s - \omega_c t)%
+u(\omega_c|\mathbf{d}|/c_s + \omega_c t)%
\Big]\,,
\end{equation}
where $u(\xi)=\cos[n\arctan(\xi)]/(1+\xi^2)^{n/2}$. The above
expression is a superposition of two waves propagating in opposite
directions. In the Ohmic case, $n=1$, the shape of the wave is
simply $u(\xi)=1/(1+\xi^2)$.

In Figure~\ref{fig4ad.eps}, we present the amplitude of $H_F(t)$,
defined via $H_F(t)={\cal H}_F\sigma _x^1 \sigma _x^2$, as well as
the sum of $H_\ind{int}$ and $H_F(t)$, for $n=1$. One can observe
that the ``onset wave'' of considerable amplitude and of shape
$u(\xi)$ propagates once between the qubits, ``switching on'' the
interaction. It does not affect the qubits once the interaction
has set in. One can also see in Figure~\ref{fig4ad.eps}, as well
as from (\ref{Eq:OS:H_F}), that the interaction between the qubits
is delayed by the time $\sim |\mathbf{d}|/c_s$, which is required
for the mediating \red{(virtual)} bosons to carry the response to the other qubit.
At the same time, relatively slow decay of the initial correction
$F(t)$ may necessitate a discussion regarding the meaning of the
``coherent'' large-time induced interaction (when the noise
effects have also fully set in). We will offer additional comments
in the concluding section.

\subsection{Initial Dynamics of the Density Matrix Elements and Entanglement}\label{ssec:DM}

Let us now take the entire solution
(\ref{Eq:OS:AdiabaticSolution}) which includes both the induced
interaction discussed above and the noise
(\ref{Eq:OS:DecoherenceFunction-ReL}). In the exact solution of
the short-time model, the bath is assumed to be thermalized
only initially. At
large enough times the externally induced equilibration of the bath should be considered.
We will account for this later when discussing the
perturbative Markovian approach. Here we continue our investigation of
the short-time model which assumes that the two qubits coupled with
the (noninteracting with each other) bath modes constitute a closed quantum system.

To quantify the interplay of the effects of the induced
interaction vs.\ noise, we evaluate the
concurrence,\cite{Wootters1,Wootters2} which measures the
entanglement of the spin system and is monotonically related to
the entanglement of formation.\cite{Bennett,Vedral} For a mixed
state of two qubits, $\rho_S$, we first define the spin-flipped
state, $ \tilde \rho _S = \sigma^1_y \sigma^2_y \, \rho^*_S \,
\sigma^1_y \sigma^2_y $, and then the Hermitian operator
$R=\sqrt{\sqrt{\rho_S}\tilde\rho_S\sqrt{\rho_S}}$, with
eigenvalues $\lambda_{i=1,2,3,4}$. The concurrence is then
given\cite{Wootters2} by
\begin{equation}\label{Eq:DM:concurence}
C\left( {\rho _S } \right) = \max \Big\{ {0,2\mathop {\max
}\limits_i \lambda _i  - \sum\limits_{j = 1}^4 {\lambda _j } }
\Big\}\,.
\end{equation}

%
\Fig{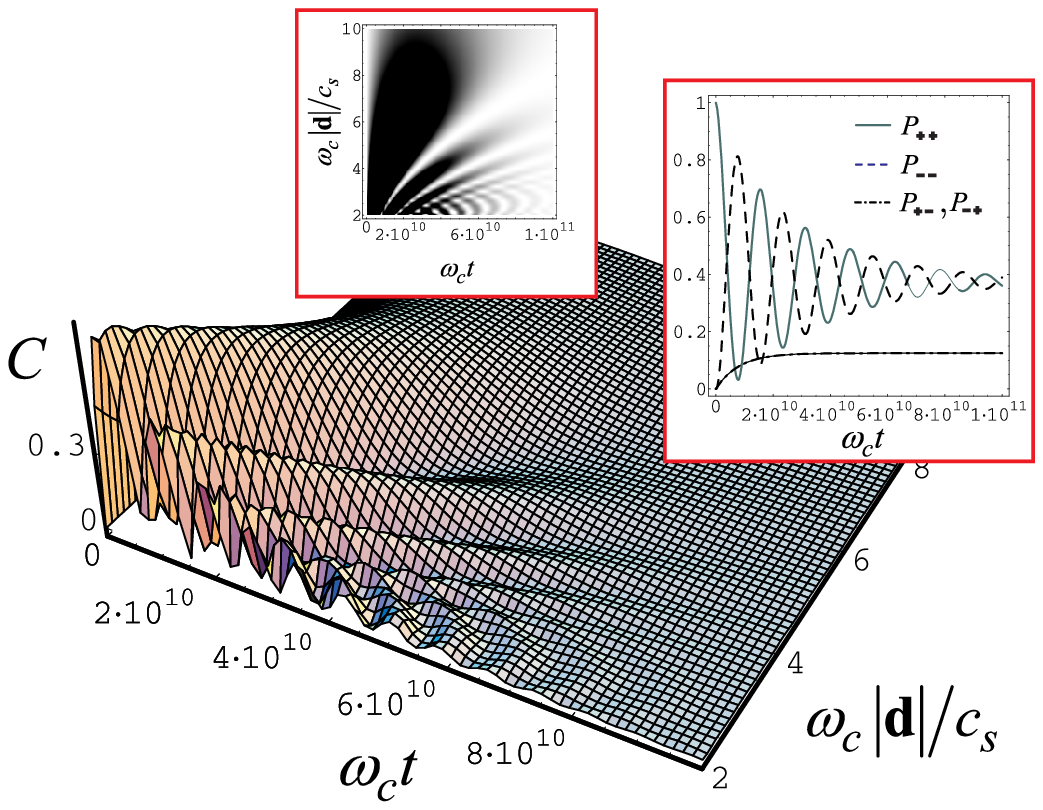}{ Development of the concurrence as a function
of time, calculated with $\alpha_1^z=0.5\cdot 10^{-7}$ and
$k_\textrm{B}T/\omega_c=0.5\cdot 10^{-2}$, which corresponds to
the magnetic field $H_z = 0.53\,$T and temperatures $T=0.34\,$K.
The left inset shows the distribution of the concurrence in the
$|\mathbf{d}|$-$t$ plane. The right inset presents the dynamics of
the diagonal density matrix elements $P_{++} \equiv \langle ++ |
\rho_S(t) | ++ \rangle $, etc., on the same time scale.}
Considering the Si channel geometry for the interaction, introduced
above as an illustrative example, we arrive at an approximately
adiabatic Hamiltonian ($\alpha_1^z \gg \alpha_1^{x,y}$) with
Ohmic-type coupling. The dynamics of the concurrence is presented
in Figure~\ref{fig_C1D_Si.eps}, and we note that the peak entanglement
can reach a sizable fraction of 1. The coupling constant $\alpha_1^z$ is quite small due to
the weakness of the spin-orbit coupling of P-impurity electrons in
Si, which results in low magnitude of the induced interaction (and
the noise due to the same environment). Nevertheless, one still
observes decaying periodic oscillations of entanglement, which indicate that
an approximately coherent dynamics can develop, due to the
bath-induced interaction, for several dynamical cycles.

To understand the dynamics of the qubit system and its
entanglement, let us continue with the analysis of the coherent
part in (\ref{Eq:OS:AdiabaticSolution}). For each spin, we define the two states
$\left|\pm\right\rangle = \left[\,\left|\uparrow\right\rangle \pm
\left|\downarrow\right\rangle \right]/{\sqrt 2}$.
After the interaction,
$H_\ind{int}$, sets in (note the time scales in
Figures~\ref{fig4ad.eps} and \ref{fig_C1D_Si.eps}), it will split
the system energies into two degenerate pairs $E_0 = E_1=-{\cal
H}_\ind{int}$ and $E_2 = E_3={\cal H}_\ind{int}$. The wave
function is then
$|\psi(t)\rangle=\exp(-iH_\ind{int}t)|\psi(0)\rangle$. For the
initial ``++'' state, $|\psi(0)\rangle=\left|++\right\rangle $, it develops as
$\left|\psi(t)\right\rangle= \left|++\right\rangle \cos {\cal
H}_\ind{int}t + \left|--\right\rangle i\sin {\cal H}_\ind{int}t$,
where $H_\ind{int} = {\cal H}_\ind{int}\sigma^1_m\sigma^2_m$. One
can easily check that at times $t_E=\pi/4{\cal
H}_\ind{int},3\pi/4{\cal H}_\ind{int},\ldots$, maximally entangled
(Bell) states are obtained, while at times $t_0=0,\pi/2{\cal
H}_\ind{int},\pi/{\cal H}_\ind{int},\ldots$, the entanglement
vanishes; these special times can also be seen in Figure~\ref{fig_C1D_Si.eps}.

The coherent dynamics just described is only approximate, because
the bath also induces decoherence that enters via
(\ref{Eq:OS:DecoherenceFunction-ReL}). The result for the
entanglement is that the decaying envelope function is
superimposed on the coherent oscillations described above. The
magnitudes of the first and subsequent peaks of the concurrence
are determined only by this function. As temperature increases,
the envelope decays faster resulting in lower values of the
concurrence.

Note also that the non-monotonic behavior of the entanglement is
possible only provided that the initial state is not trivial with
respect to the induced interaction, otherwise only the phase
factor is developed. For example, taking the initial state
$\left[\,\left|-+\right\rangle + \left|+-\right\rangle \right]/\sqrt{2}$, in our
case would only lead to the destruction of entanglement, i.e., to
a monotonically decreasing concurrence, similar to results of
other studies.\cite{Eberly1,Eberly2}

\red{For the model that allows the exact solution, i.e., for $H_S=0$,
one can notice that there is no relaxation by energy transfer
between the system and bath. The exponentials in
(\ref{Eq:OS:AdiabaticSolution}), with
(\ref{Eq:OS:DecoherenceFunction-ReL}), suppress only the
off-diagonal matrix elements, i.e., those with
$\lambda\neq\lambda'$. It happens, however, that at large times
the $\mathbf{d}$-dependence (the cosine term in
$\mathrm{Re}\,\frak{L}_{\lambda\lambda'}$) is not important in
(\ref{Eq:OS:DecoherenceFunction-ReL}). Therefore, one can show that
$\textrm{Re}\,\frak{L}_{\lambda\lambda'}(t\rightarrow\infty)$
vanishes for certain combinations of $\lambda\neq\lambda'$. Specifically, the limiting
$t\to \infty$ density matrix for our initial state
($\left|++\right\rangle$) retains some non-diagonal elements,
\begin{equation}\label{Eq:DM:roLimit}
\rho(t\rightarrow\infty)\rightarrow\frac{1} {8}\left( {
{\begin{array}{*{20}c}
   3 & 0 & 0 & { - 1}  \\
   0 & 1 & 1 & 0  \\
   0 & 1 & 1 & 0  \\
   { - 1} & 0 & 0 & 3  \\
 \end{array} }} \right)\, .
\end{equation}
The basis states here are $\left|++\right\rangle$,
$\left|+-\right\rangle$, $\left|-+\right\rangle$, and
$\left|--\right\rangle$. The significance of this and
similar\cite{Braun} results is in the fact that in the model with
$H_S=0$ and non-rethermalizing bath not all the off-diagonal
matrix elements need be suppressed by decoherence, even though the
concurrence of (\ref{Eq:DM:roLimit}) is zero.}

\section{INTERPLAY BETWEEN THE INDUCED INTERACTION AND QUANTUM NOISE AT LARGER TIMES}\label{sec:LT}

\subsection{Master-Equation Approach}\label{ssec:ME}

In this section we present the expressions for the induced
interaction and also for the noise effects due to the bosonic
environment, calculated perturbatively to the second order in the
spin-boson interaction, and with the assumption that the
environment is constantly reset to thermal.\cite{STPb}

The dynamics of the system can be described by the equation for
the density matrix,
\begin{equation}\label{Eq:ME:Liuvolle}
i\dot \rho (t) = [H,\rho (t)] \, .
\end{equation}
In order to trace over the bath variables, we carry out  the
second-order perturbative expansion. This dynamical description is
supplemented by the set of Markovian
assumptions,\cite{Leggett,VKampen,Louisell,Blum} one of which
invokes resetting the bath to thermal equilibrium, at temperature
$T$, after each infinitesimal time step, as well as at time $t=0$,
see (\ref{Eq:M:IC}), thereby decoupling the qubit system from the
environment.\cite{Leggett,VKampen} This is a physical assumption
appropriate for all but the shortest time scales of the system
dynamics.\cite{Privman,Tolkunov2,Solenov} It can also be viewed as
a means to phenomenologically account in part for the
randomization of the bath modes due to their interactions with
each other (anharmonicity) in real systems. This leads to the
master equation for the reduced density matrix of the qubits, $\rho
_S (t) = Tr_B \rho (t) $,
\begin{equation}\label{Eq:ME:MME}
i\dot \rho _S (t) = [H_S ,\rho _S (t)]
- i \!\int\limits_0^\infty {dt'Tr_B [H_{SB} ,[H_{SB} (t' - t),\rho
_B \rho _S (t)]]} \, ,
\end{equation}
where $H_{SB} (\tau)=e^{i(H_B+H_S)\tau}H_{SB}e^{-i(H_B+H_S)\tau}$.
Analyzing the structure of the integrand in (\ref{Eq:ME:MME}),
after lengthy calculations\cite{STPb} one can obtain the equation
with explicitly separated coherent and noise contributions,
\begin{equation}\label{Eq:ME:MRateEq}
i\dot \rho _S (t) = [H_\textrm{eff} ,\rho _S (t)] + i{\hat M}\rho
_S (t)\, .
\end{equation}
The effective coherent Hamiltonian $H_\textrm{eff}$ is
\begin{equation}
H_\textrm{eff}\!\!\!=  H_S  +\!\!\!\! \sum\limits_{m=x,y,z}
\!\!\! {2\chi _c^m ({\mathbf{d}})\sigma _m^1 \sigma _m^2 }  - \chi
_s^x ({\mathbf{d}})\left( {\sigma _x^1 \sigma _y^2  + \sigma _x^2
\sigma _y^1 } \right)
+ \chi _s^y ({\mathbf{d}})\left( {\sigma _y^1 \sigma _x^2  +
\sigma _y^2 \sigma _x^1 } \right) -
\left[\eta_s^x(0)+\eta_s^y(0)\right]
\left({\sigma_z^1+\sigma_z^2}\right)\, . \label{Eq:ME:H-eff}
\end{equation}
The expressions for the amplitudes $\chi_c^m(\mathbf{d})$,
$\chi_s^m(\mathbf{d})$, $\eta_c^m(\mathbf{d})$, and
$\eta_s^m(\mathbf{d})$ will be given bellow. The first three terms
following $H_S$ constitute the interaction between the two spins.
We will argue below that the leading induced exchange interaction
is given by the first added term, proportional to
$\chi_c^m(\mathbf{d})$. The last term gives the qubit Lamb shifts.

The otherwise cumbersome expression for the noise terms can be
represented concisely by introducing the noise superoperator $\hat
M$, which involves single-qubit contributions, which are usually
dominant, as well as two-qubits terms,
\begin{equation}\label{Eq:ME:M-sums}
{\hat M} = \sum\limits_{m,i} \Big[ {\hat M}_m^i (0) +\sum_{j \ne
i} {\hat M}_m^{ij} (\mathbf{d} ) \Big] \,,
\end{equation}
where the summations are over the components, $m=x,y,z$, and the
qubits, $i,j=1,2$. One can define the superoperators ${\hat L}_a
(O_1)O_2=\{O_1,O_2\}$, ${\hat L}(O_1,O_2)O_3=O_1O_3O_2$, and
${\hat L}_\pm(O_1,O_2)={\hat L}(O_1,O_2)\pm {\hat L}(O_2,O_1)$,
to write
\begin{equation}\label{Eq:ME:M-cross}
{\hat M}_m^{ij}(\mathbf{d})= \eta_c^m(\mathbf{d}) \left[{2{\hat
L}(\sigma^i_m,\sigma^j_m) - {\hat L}_a(\sigma^i_m\sigma^j_m)}
\right]
+ \eta_s^m(\mathbf{d})\left[{{\hat
L}_+(\sigma^i_m,\varsigma^j_{m})-{\hat L}_a(\sigma^i_m
\varsigma^j_{m} )}\right]
- i\chi_s^m(\mathbf{d}){\hat L}_-(\sigma^i_m ,\varsigma^j_{m})\, ,
\end{equation}
where we denote $\varsigma^j_{m} = \frac{i}{2}[\sigma_z^j ,\sigma
_m^j]$, and
\begin{equation}\label{Eq:ME:M-local}
{\hat M}_m^j(0)= \eta _c^m (0)\left[{2{\hat L}(\sigma _m^j,\sigma
_m^j) -{\hat L}_a(\sigma _m^j \sigma _m^j )}\right]
 + \eta _s^m(0){\hat L}_+(\sigma_m^j ,\varsigma^j_{m})
- i\chi _s^m (0)\left[{{\hat
L}_-(\sigma_m^j,\varsigma^j_{m})+{\hat L}_a(\sigma_m^j
\varsigma^j_{m})}\right]\,.
\end{equation}

The amplitudes in (\ref{Eq:ME:H-eff}), (\ref{Eq:ME:M-cross}), and
(\ref{Eq:ME:M-local}), calculated for the interaction defined in
(\ref{Eq:M:H_SB})-(\ref{Eq:M:X_jm}), are
\begin{equation}\label{Eq:ME:he-c}
\chi _c^m ({\mathbf{d}}) =  - \sum\limits_\mathbf{\xi}
{\int\limits_{ - \infty }^\infty  {\frac{{Vd{\mathbf{k}}}}
{{\left( {2\pi } \right)^3 }}} } \left|
{g_{{\mathbf{k}},\mathbf{\xi}}^m } \right|^2 \frac{{\omega
_{{\mathbf{k}},\mathbf{\xi}} \cos \left( {{\mathbf{k}} \cdot
{\mathbf{d}}} \right)}} {{\omega _{{\mathbf{k}},\mathbf{\xi}}^2  -
\Delta ^2 (1 - \delta _{m,z} )}}\,,
\end{equation}
\begin{equation}\label{Eq:ME:eta-c}
\eta _c^m ({\mathbf{d}}) = \frac{\pi}{2}\sum\limits_\mathbf{\xi}
{\int\limits_{ - \infty }^\infty  {\frac{{Vd{\mathbf{k}}}}
{{\left( {2\pi } \right)^3 }}} } \left|
{g_{{\mathbf{k}},\mathbf{\xi}}^m } \right|^2 \coth \frac{{\omega
_{{\mathbf{k}},\mathbf{\xi}} }} {{2k_\textrm{B}T}}\cos \left(
{{\mathbf{k}} \cdot {\mathbf{d}}} \right)
\sum\limits_{q = \pm 1} {\delta \left( {\omega
_{{\mathbf{k}},\mathbf{\xi}} + (1 - \delta _{m,z} )q\Delta }
\right)}\,,
\end{equation}
\begin{equation}\label{Eq:ME:he-s}
\chi _s^m ({\mathbf{d}}) =  - \left( {1 - \delta _{m,z} }
\right)\sum\limits_\mathbf{\xi} {\int\limits_{ - \infty }^\infty
{\frac{{Vd{\mathbf{k}}}} {{\left( {2\pi } \right)^3 }}} } \left|
{g_{{\mathbf{k}},\mathbf{\xi}}^m } \right|^2 \cos \left(
{{\mathbf{k}} \cdot {\mathbf{d}}} \right)
\sum\limits_{q =  \pm 1} {\frac{{\pi s}} {2}\delta \left(
{\omega _{{\mathbf{k}},\mathbf{\xi}}  + q\Delta } \right)}\, ,
\end{equation}
\begin{equation}\label{Eq:ME:eta-s}
\eta _s^m ({\mathbf{d}}) = \left( {1 - \delta _{m,z} } \right)
\sum\limits_\mathbf{\xi} {\int\limits_{ - \infty }^\infty
{\frac{{Vd{\mathbf{k}}}} {{\left( {2\pi } \right)^3 }}} } \left|
{g_{{\mathbf{k}},\mathbf{\xi}}^m } \right|^2 \coth \frac{{\omega
_{{\mathbf{k}},\mathbf{\xi}} }} {{2k_\textrm{B}T}}\frac{{\Delta
\cos \left( {{\mathbf{k}} \cdot {\mathbf{d}}} \right)}} {{\omega
_{{\mathbf{k}},\mathbf{\xi}}^2  - \Delta ^2 }}\,.
\end{equation}
Here the principal values of integrals are assumed.

Note that $\chi _c^m ({\mathbf{d}})$ appears only in the induced
interaction Hamiltonian in (\ref{Eq:ME:H-eff}), whereas $\eta
_c^m ({\mathbf{d}})$, $\chi _s^m ({\mathbf{d}})$, and $\eta _s^m
({\mathbf{d}})$ enter both the interaction and noise terms.
Therefore, in order to establish that the induced interaction can
be significant for some time scales, we have to demonstrate that
$\chi _c^m ({\mathbf{d}})$ can have a much larger magnitude than
the maximum of the magnitudes of $\eta _c^m ({\mathbf{d}})$, $\chi
_s^m ({\mathbf{d}})$, and $\eta _s^m ({\mathbf{d}})$. The third
and fourth terms in the expression for the interaction
(\ref{Eq:ME:H-eff}) are comparable to the noise and therefore have
no significant contribution to the coherent dynamics.

\subsection{Qubit Coupling in a 1D Channel}\label{ssec:1D}

It is instructive to start with the simple 1D example, leaving the
derivations for higher dimensions to the next subsection. The 1D
geometry can actually be natural for certain ion-trap
quantum-computing schemes, in which ions in a chain are subject to
Coulomb interaction, developing a variety of phonon-mode lattice
vibrations.\cite{Leibfried,Marquet,Porras}

In our case we allow the phonons to propagate in a single
direction, along $\mathbf{d}$, so that ${\mathbf{k}} \cdot
{\mathbf{d}} = k\left| {\mathbf{d}} \right|$. Here, for
definiteness, we also assume the linear dispersion, $\omega _k  =
c_s k$. Furthermore, we will ignore the polarization index in the
coupling constants: $g_{{\mathbf{k}},\mathbf{\xi}}^m  \to g^m
(\omega )$.

The induced interaction and noise terms depend on the amplitudes
$\chi _c^m ({\mathbf{d}})$, $\eta _c^m ({\mathbf{d}})$, $\chi _s^m
({\mathbf{d}})$, and $\eta _s^m ({\mathbf{d}})$, two of which can
be evaluated explicitly for the 1D case, because of the
$\delta$-functions \red{in (\ref{Eq:ME:eta-c}), (\ref{Eq:ME:he-s}).}
However, to derive an explicit expression for $\chi _c^m
({\mathbf{d}})$ and $\eta _s^m ({\mathbf{d}})$, one needs to
specify the $\omega$-dependence in $\left| {g^m (\omega )}
\right|^2$. For the sake of simplicity, in this section we
approximate $|g^m (\omega )|^2$ by a linear function with
superimposed exponential cutoff. For a constant 1D density of
states $V/2\pi c_s$, this is a variant of the
Ohmic-dissipation\cite{Leggett} model, i.e., (\ref{Eq:M:DOS}) with
$n=1$ (the case when $n>1$ is considered in the next subsection).

In most practical applications, we expect that $\Delta \ll
\omega_c$. With this assumption, we obtain
\begin{equation}\label{Eq:1D:he-c-explicit-1D}
\chi _c^m ({\mathbf{d}}) = \frac{{\alpha _1^m \omega _c }} {{1 +
\left( {{{\omega _c \left| {\mathbf{d}} \right|} \mathord{\left/
 {\vphantom {{\omega _c \left| {\mathbf{d}} \right|} {c_s }}} \right.
 \kern-\nulldelimiterspace} {c_s }}} \right)^2 }}\,,
\end{equation}
\begin{equation}\label{Eq:1D:he-s-explicit-1D}
\chi _s^m ({\mathbf{d}}) = \alpha _1^m \omega _c \frac{\pi }
{2}\left( {1 - \delta _{m,z} } \right)\frac{\Delta } {{\omega _c
}}\cos \frac{{\Delta \left| {\mathbf{d}} \right|}} {{c_s }}\,,
\end{equation}
\begin{equation}\label{Eq:1D:eta-c-explicit-1D}
\eta _c^m ({\mathbf{d}}) = \alpha _1^m \omega _c \frac{\pi }
{2}\left( {1 - \delta _{m,z} } \right)\frac{\Delta } {{\omega _c
}}\coth \frac{\Delta } {{2k_\textrm{B}T}}\cos \frac{{\Delta \left|
{\mathbf{d}} \right|}} {{c_s }}\,.
\end{equation}
The expression for $\eta _s({\mathbf{d}})$ could not be obtained
in closed form. However, numerical estimates suggest that  $\eta
_s({\mathbf{d}})$ is comparable to $\eta _c^m ({\mathbf{d}})$. At
short spin separations $\mathbf{d}$, $\eta _s({\mathbf{d}})$ is
approximately bounded by $- \alpha_1^m \Delta
\ln\frac{\Delta}{\omega_c} \exp(-\frac{{\Delta \left| {\mathbf{d}}
\right|}} {{c_s }})$, while at lager distances it may be
approximated by $\alpha_1^m \Delta\frac{\pi}{2}\sin \frac{{\Delta
\left| {\mathbf{d}} \right|}} {{c_s }}
\coth\frac{\Delta}{2k_\textrm{B}T}$. The level of noise may be
estimated by considering the quantity
\begin{equation}\label{Eq:1D:M-explicit-1D}
{\cal M} = \max_{|\mathbf{d}|} \left|{\eta _c^m(\mathbf{d}), \chi
_s^m(\mathbf{d}), \eta _s^m(\mathbf{d})} \right|\,.
\end{equation}

The interaction Hamiltonian takes the form
\begin{equation}\label{Eq:1D:H-int-1D}
H_{\operatorname{int} }  =  - \frac{2} {{1 + \left( {{{\omega _c
\left| {\mathbf{d}} \right|}/{c_s }}} \right)^2 }}
 \sum\limits_{m = x,y,z} {\alpha _1^m \omega _c \sigma _m^1 \sigma
 _m^2}\,.
\end{equation}
This induced interaction is temperature independent. It is
long-range and decays as a power law for large $\mathbf{d}$. Note
that the expression (\ref{Eq:1D:H-int-1D}) is the same as the
induced interaction obtained from the short-time model
(\ref{Eq:OS:H-int}), provided the proper spin-phonon interaction
component is considered.

If the noise term were not present, the spin system would be
governed by the Hamiltonian $H_S  + H_{\textrm{int}}$. To be
specific, let us analyze the spectrum, for instance, for $\alpha
_1^x  = \alpha _1^y$, $\alpha _1^z  = 0$. The two-qubit
states would consist of the singlet
$(\left|{\uparrow\downarrow}\right\rangle-
\left|{\downarrow\uparrow}\right\rangle)/\sqrt{2}$ and the split
triplet $\left|{\uparrow\uparrow}\right\rangle$,
$(\left|{\uparrow\downarrow}\right\rangle+
\left|{\downarrow\uparrow}\right\rangle)/\sqrt{2}$, and
$\left|{\downarrow\downarrow}\right\rangle$, with energies
$E_2=-4\chi _c^x$, $E_0=-\Delta$, $E_1=4\chi _c^x$, $E_3=\Delta$,
respectively. The energy gap $|E_1-E_2|$ between the two entangled
states is defined by $4\chi _c^x$ ($=4\chi _c^y$). In the presence
of noise, the oscillatory, approximately coherent evolution of the
spins can be observed over several oscillation cycles provided
that $2\alpha_1^m \omega_c/[1 +(\omega_c |\mathbf{d}|/c_s)^2]>
{\cal M}$. The energy levels will acquire effective width due to
quantum noise, of order $\eta_c^m(\mathbf{d})$.

\subsection{Boson-Mediated Induced Qubit-Qubit Interaction in General Dimensions}\label{ssec:GD}

Let us generalize the results of the previous subsections, where
we considered the 1D case with Ohmic dissipation within the
Markovian approach. In the general case, let us consider the
Markovian model and, again, assume that $\Delta/\omega_c$ is
small. We will also assume that the absolute square of the $m$th
component of the spin-boson coupling, when multiplied by the
density of states, can be modeled by
$\alpha_{n_m}^m{}\omega^{n_m}\exp(-\omega/\omega_c)$; see
(\ref{Eq:M:DOS}). The integration in (\ref{Eq:ME:he-c}) can then
be carried out in closed form for any $n_m=1,2,\ldots\,\,$. The
induced interaction (\ref{Eq:1D:H-int-1D}) is thus generalized to
\begin{equation}\label{Eq:GD:H-int}
H_\textrm{int}=-\!\!\!\sum\limits_{m =
x,y,z}{\alpha_{n_m}^m\omega_c^{n_m }\sigma_m^1\sigma _m^2}
\frac{{2\Gamma (n_m
)\textrm{Re}\left({1+i\omega_c\left|{\mathbf{d}}\right|/c_s}\right)^{n_m}}}
{{\left[{1+\left({\omega_c\left|{\mathbf{d}}\right|/c_s}\right)^2}\right]^{n_m}}}\,.
\end{equation}

With the appropriate choice of parameters, the result for the
induced interaction, but not for the noise, coincides with the
expression for $H_\textrm{int}$ obtained within the short-time
model. The effective interaction has the large-distance asymptotic
behavior $\left|{\mathbf{d}}\right|^{-n_m}$, for even $n_m$, and
$\left|{\mathbf{d}}\right|^{-n_m-1}$, for odd $n_m$.

In higher dimensions the structure of
$g_{{\mathbf{k}},\mathbf{\xi}}^m$ in the $\textbf{k}$-space
becomes important. Provided $\omega _{{\mathbf{k}},\mathbf{\xi}}$
is nearly isotropic, the integrals in
(\ref{Eq:ME:he-c})-(\ref{Eq:ME:eta-s}) will include (in 3D) a
factor
$\int_0^{2\pi}{d\varphi}\int_0^\pi{d\theta\sin\theta}\left|{
g_{k\theta\varphi,\mathbf{\xi}}^m}\right|^2\cos\left({k\left|{\mathbf{d}}
\right|\cos\theta}\right)$, which can be written as
$[f^m_1(\omega,k|\mathbf{d}|)-f^m_2(\omega,
k|\mathbf{d}|)\partial/\partial|\mathbf{d}|]\cos (k|\mathbf{d}|)$,
\red{see Eqs.~(B1)-(B4) of Ref.~2 
for details.} When the
dependence of $f^m_1,f^m_2$ on $k|\mathbf{d}|$ is negligible, the
interaction is simply $H_\textrm{int}\rightarrow
H_\textrm{int}|_{\{n_m\}\rightarrow
a}-(\partial/\partial|\mathbf{d}|)H_\textrm{int}|_{\{n_m\}\rightarrow
b}$, where $a$ and $b$ are sets of three integers representing the
$\omega$-dependence of $f_1^m(\omega,k|\mathbf{d}|)$ and
$f_2^m(\omega,k|\mathbf{d}|)$. Otherwise, a more complicated
dependence on $|\mathbf{d}|$ is expected. The noise superoperator
can be treated similarly.

\subsection{Induced Interaction vs.\ Noise in the Semiconductor Impurity Qubit Model}\label{ssec:3D}

Let us now proceed to an example of the semiconductor qubit model
in the bulk material, which gives usual 3D geometry for phonon
propagation. We consider spins of P-impurity donor electrons as
qubits. Dilute P-impurities are embedded in the matrix of a Si or
Ge crystal kept at a sufficiently low temperature such that the
outer electrons remain bound. The two impurities placed next to
each other at distances of about 10-30$\,$nm would constitute a
two qubit register. The spins would then be interacting indirectly
via the spin-orbit coupling and lattice deformation (phonons) as
well as directly via the dipole-dipole electromagnetic coupling. In
this section we will estimate and compare both. We will also
present the dynamics due to the coherent vs.\ noise induced
features and their interplay.

Let us consider for simplicity only the LA phonon branch, ${\mathbf{\xi }} \to
{\mathbf{k}}/\left| {\mathbf{k}} \right|$, and assume an isotropic
dispersion
$\omega_{\mathbf{k},\mathbf{\xi}}=c_s\left|\mathbf{k}\right|$. The
expression for the coupling constants is then
\begin{equation}\label{Eq:3D:gk-3D}
g_{{\mathbf{k}},{\mathbf{\xi }}}^m  = D_m \frac{{k_z k_m }}
{{\left( {1 + a_\textrm{B}^2 k^2 } \right)^2 }}\sqrt {\frac{\hbar
} {{2\rho Vc_s k^3 }}}\,.
\end{equation}
One can show\cite{STPb} that the cross terms, with $m \neq n$, of
the correlation functions
$Tr_B \left[ X_m^j X_n^i (t)\rho_B \right]$, depend
on the combination $g_{\mathbf{k},
\mathbf{\xi}}^m(g_{\mathbf{k}, \mathbf{\xi}}^n)^*$, which is
always an odd function of one of the projections of the wave
vector. Thus, all the non-diagonal terms vanish.

Integrating (\ref{Eq:ME:he-c}) and (\ref{Eq:ME:eta-c}) with
(\ref{Eq:3D:gk-3D}), one can demonstrate\cite{STPb} that
decoherence is dominated by the individual noise terms for each
spin, with the typical amplitude
\begin{equation}\label{Eq:3D:eta-c-xy}
\eta _c^{x,y} (0) = C_\textrm{B} \frac{{2\pi ^2 }}
{{15}}\frac{b^3}{\left(
{1+b^2}\right)^4}\coth\frac{\Delta}{2k_\textrm{B}T}\,,
\end{equation}
where $b = \Delta a_{\rm B} /c_s$ and $C_\textrm{B}={\rm B}^2\mu_{\textrm{B}}^2
H_z^2/\left({16\pi^3\rho \hbar a_{\rm B}^3 c_s^2}\right)$. The
interaction amplitude $\chi _c^m ({\mathbf{d}})$ and, therefore,
the induced spin-spin interaction, have inverse-square power-law
asymptotic form for the $x$ and $y$ spin components, with a
superimposed oscillation, and inverse-fifth-power-law decay for
the $z$ spin components,
\begin{eqnarray}\label{Eq:3D:H-int-asymptotic}
H_\textrm{int}&=&\sum\nolimits_m {2\chi _c^m ({\mathbf{d}})\sigma
_m^1 \sigma _m^2 }
\\ \nonumber
&\xrightarrow{{r\gg1}}& - 4\pi ^2 C_\textrm{B} \frac{{2b}}
{{\left( {1 + b^2 } \right)^4 }}\frac{{\sin br}} {{r^2 }}\left(
{\sigma _x^1 \sigma _x^2  + \sigma _y^1 \sigma _y^2 } \right) +
384\pi ^2 C_\textrm{A} \frac{2} {{r^5 }}\sigma _z^1 \sigma _z^2
\\ \nonumber
&\xrightarrow{{r\ll 1}}& \left[ - 4 \pi^2 C_\mathrm{B}
\frac{1+9b^2-9b^4+b^6}{120 (1+b^2)^4} + {\cal O}(r^2)
\right]\left( {\sigma _x^1 \sigma _x^2  + \sigma _y^1 \sigma _y^2
} \right) + \left[ - C_\mathrm{A}\frac{\pi^2}{10} +{\cal O}(r^2)
\right] \sigma _z^1 \sigma _z^2 \,.
\end{eqnarray}
Here $r=|\mathbf{d}|/a_\textrm{B}$ and
$C_\textrm{A}=\textrm{A}^2C_\textrm{B}/\textrm{B}^2$. At small
distances the interaction is regular and the amplitudes converge
to constant values, see Figure~\ref{figHM.eps}. The complete
expressions for $\chi_c^m(\mathbf{d})$ and $\eta_c^m(\mathbf{d})$
are also available.\cite{STPb}

In Figure~\ref{figHM.eps}, we plot the amplitudes of the induced
spin-spin interaction, which has the asymptotic behavior
(\ref{Eq:3D:H-int-asymptotic}), and noise for different values of
the spin-spin separation and $b$, for electron impurity spins in
3D Si-Ge type structures. The value of $b$ can be controlled via
the applied magnetic field, $b=\mu_\textrm{B}H_z g^*
a_\textrm{B}/c_s$. The temperature dependence of the noise is
insignificant provided $2k_{\textrm{B}}T/\Delta\ll 1$.

As mentioned earlier, the obtained interaction
(\ref{Eq:3D:H-int-asymptotic}) is always accompanied by noise
coming from the same source, as well as by the direct
interactions of the spins. When the electron wave
functions overlap is negligible, the dominant direct interaction
will be the electromagnetic dipole-dipole one,
\begin{equation}\label{Eq:3D:EM-spin-spin-interaction}
H_\textrm{EM}(\mathbf{d})=\frac{\mu_0 \mu _\textrm{B}^2}{4\pi}
\frac{\sigma _x^1 \sigma _x^2  + \sigma _y^1 \sigma _y^2  -
2\sigma _z^1 \sigma _z^2}{|\mathbf{d}|^3}\,.
\end{equation}
The comparison of the two interactions and noise is shown in
Figure~\ref{figGeHEM.eps}. We plot the magnitude of the effective
induced interaction (\ref{Eq:3D:H-int-asymptotic}), the
electromagnetic interaction
(\ref{Eq:3D:EM-spin-spin-interaction}), measured by ${\cal
H}_\textrm{EM}\equiv{\mu_0 \mu _\textrm{B}^2}/{4\pi
|\mathbf{d}|^3}$, and a measure of the level of noise, for P-donor
electron spins in Ge. In the plot the coupling constants for
P-donors in Ge were taken as
$C_\textrm{B}=1.3\times{}10^7\,$s$^{-1}$ and
$C_\textrm{A}\approx0$, see Sec.~\ref{sec:M}. It transpires that
the induced interaction can be considerable as compared to the
electromagnetic spin-spin coupling. However the overall
coherence-to-noise ratio is quite poor for Ge. In Si, the level of
noise is lower as compared to the induced interaction. It is
dominated primarily by the adiabatic ($\sigma_z^{1,2}$) term.
However, the overall amplitude of the induced terms compares less
favorably with the electromagnetic coupling.
\Fig{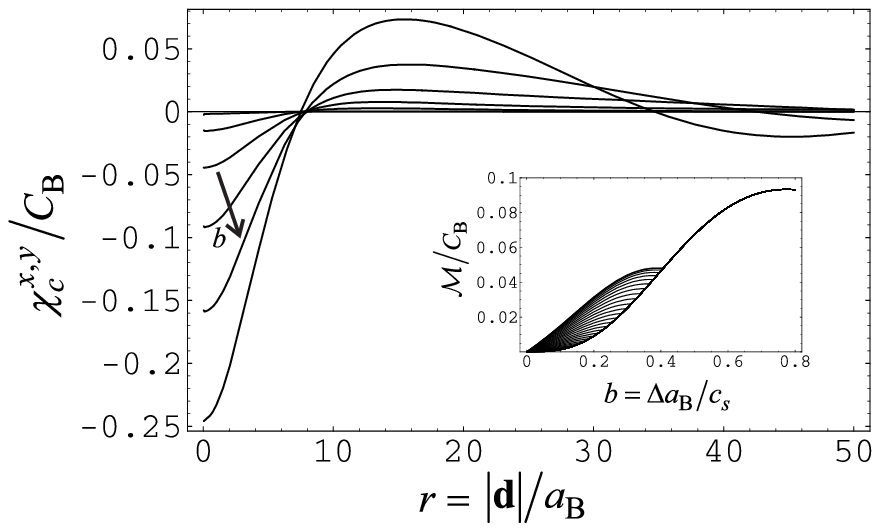}{The magnitude of the induced spin-spin interaction
for a 3D Si-Ge type structure: the dominant interaction amplitude,
which is the same for the $x$ and $y$ spin components, is shown.
The arrow indicates increasing $b$ values for the curves shown,
with $b=\Delta a_{\rm B} /c_s=0.03$, $0.09$, $0.15$, $0.21$, $0.27$, $0.33$. The inset
estimates the level of the noise (for $2k_{\textrm{B}}Ta_{\rm B}/c_s=0.01$): the
bottom curve is $\eta_c^{m}(0)$, with $m=x,y$. For $b \leq 0.4$,
the amplitude $\eta_s^{m}(\mathbf{d})$ can be comparable, and its
values, calculated numerically for $0\leq r\leq 50$, are shown as
long as they exceed $\eta_c^{m}(0)$, with the top curve
corresponding to the maximum value, at $r=0$.}
\Fig{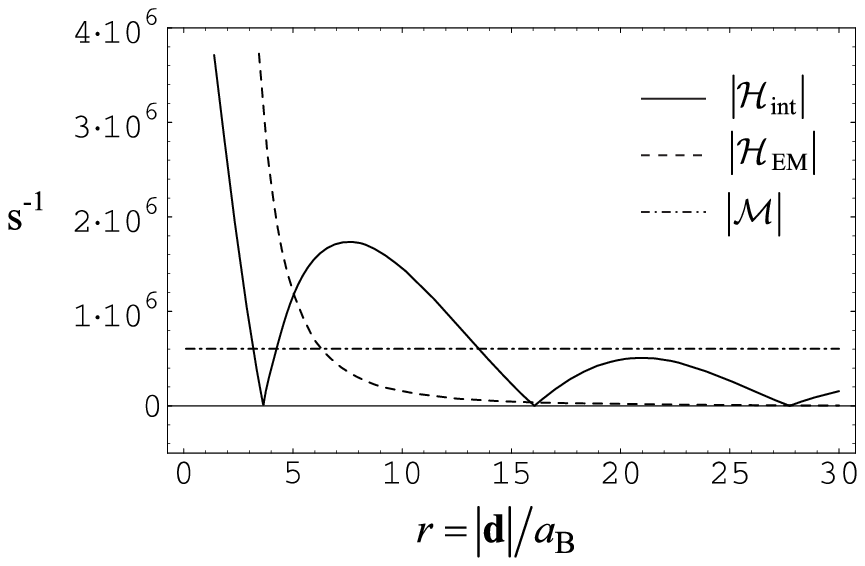}{The magnitudes, measured in units of s$^{-1}$,
of the induced spin-spin interaction, the dipole-dipole coupling strength,
and the level of noise for P-impurity electron spins in Ge. Here
$H_z=3 \times 10^4\,$G, and low temperature,
$2k_\textrm{B}T/\Delta \ll 1$, was assumed.}

Let us estimate the development of entanglement for the P-in-Si
case, considering only the bath-induced effects.
Taking $\left|++\right\rangle$ as the initial state, one expects
entanglement to develop due to the ${\cal H}_\ind{int} \sigma_z^1
\sigma_z^2$ interaction term. The corresponding coupling constant
is $C_\textrm{B}=7800\,$s$^{-1}$. At distances of about
$12a_\ind{B}$ this part of the interaction can be already well
approximated by ${\cal H}_\ind{int} \sim 384\pi ^2 C_\textrm{A}
\frac{2 a_\ind{B}^5} {{|\mathbf{d}|^5 }}\sim 100\,$s$^{-1}$.
The $x$ and $y$ components of the
interaction are smaller by a factor $10^{-3}$ and will not be
considered. The pure state would then be developing as
$\left|\psi(t)\right\rangle= \left|++\right\rangle \cos {\cal
H}_\ind{int}t + \left|--\right\rangle i\sin {\cal H}_\ind{int}t$.
At times $t_E=\pi/4{\cal H}_\ind{int},3\pi/4{\cal
H}_\ind{int},\ldots$, the maximally entangled (Bell) states are
obtained, with the first occurring at $\simeq 0.8 \cdot 10^{-2}\,$s, and
then at intervals of $\simeq 1.6 \cdot 10^{-2}\,$s. The actual
entanglement is lowered by (i) the short-time (adiabatic) decoherence
term, (ii) the large-time relaxation amplitudes (rates). The
latter are due to the $x,y$ components of the qubit-bath coupling
and are of order $\sim 1\,$s$^{-1}$. As a result the first peak of the
concurrence is lowered by a factor of $\sim \exp(-0.8 \cdot 10^{-2})$,
and the subsequent peaks decrease by $\sim \exp(-1.6 \cdot 10^{-2})$ per
each cycle.

\hphantom{A}

\section{DISCUSSION OF OPEN PROBLEMS}\label{sec:discuss}

We have studied the induced indirect exchange interaction due to a
bosonic bath which also introduces quantum noise. At certain time
scales the induced two-qubit interaction is sufficiently strong to
produce significant coherent effects. This interaction can be a
factor to be considered in designs of solid state (as well as
ion-trap based) qubit registers. Even more importantly, the fact
that noise-inducing environment can also yield coherent features
in the dynamics of open quantum systems poses several interesting
fundamental challenges.

The usual approach to
thermalization\cite{VKampen,Louisell,Blum,Abragam,Slichter} has
been to assume that, for large enough times, the time evolution of
the system plus bath is not just covered by the combined
Hamiltonian, but is supplemented by the instantaneous
loss-of-memory (Markovian) approximation, which introduces
irreversibility and imposes the bath temperature on the reduced
system dynamics in the infinite-time limit, which is then
approached as the density matrix elements assume their thermal
values, according to
\begin{equation} \label{Eq:Dis:STh}
\rho _S (t) \to {{e^{ - H_S /kT} }\Big / {{{Tr}}\left( {e^{ - H_S
/kT} } \right)}} \,,
\end{equation}
with exponential relaxation (diagonal) and decoherence
(off-diagonal) rates defining the time scales $T_{1,2}$. However,
it turns out that the traditional phenomenological no-memory
approximations, yielding thermalization, the Fermi golden rule for
the transition rates, etc., assume in a way too strong a memory
loss: they erase the possible bath contributions to the coherent
part of the dynamics at shorter times, such as the Lamb shift for
a single system as well as the induced RKKY interactions for a
bi-partite system. Indeed, while relaxation leading to
(\ref{Eq:Dis:STh}) is driven by the ``on-shell'' exchanges with
the bath, it is the memory of (correlation, entanglement with) the
bath modes that drives, via virtual exchanges, the induced
interaction. Actually, the ``on-shell'' condition, imposed by the
so-called secular \red{approximation,\cite{Koppens} also} underestimates
additional decoherence at short times$\,$---$\,$the ``pure'' or
``adiabatic'' contribution to the off-diagonal
dephasing$\,$---$\,$that has thus been estimated by using other
approaches.\cite{FFPReview,Privman,Privman2,Tolkunov2,Paladino}

Perhaps the simplest way to recognize the ambiguity is to ask if
the Hamiltonian in (\ref{Eq:Dis:STh}) should have included the
bath-induced interaction terms (not shown)? Should the final
Boltzmann distribution correspond to the energy levels/basis
states of the original ``bare'' system or the one with the
RKKY-interaction/Lamb shifts, and more generally, a
bath-renormalized, ``dressed'' system?

There is presently no consistent treatment that will address in a
satisfactory way all the expected \textit{physics} of the
bath-mode effects on the dynamics. The issue is partly technical,
because we are after a \textit{tractable}, rather than just a
``foundational'' answer. It is well accepted that the emergence of
irreversibility cannot be fully treated within tractable and
calculationally convenient approaches derived directly from the
microscopic dynamics: phenomenology has to be appealed to.
However, even allowing for phenomenological solutions, \red{most of the
known tractable Markovian-approximation-involving schemes that allow for
the emergence of the indirect exchange interaction,
treat thermalization in a cavalier way, yielding typically the
noise term corresponding to $T$-dependent relaxation, but
in the strict $t\to \infty$ limit resulting
in the completely random ($T=\infty$) density matrix (proportional
to the unit operator). This then
avoids the issue of which $H_S$ should
enter in (\ref{Eq:Dis:STh}). And, as mentioned, the established
schemes that yield a more realistic, thermalized
density matrix at $t=\infty$, lose some
intermediate- and short-time dynamical effects.}

Thus, we have discussed the challenges in formulating unified
treatments that will cover all the (or just most of the
interesting) dynamical effects, over several time scales, from
short to intermediate times (for induced interaction effects and
pure decoherence) to large times (for the onset of
thermalization), while providing a tractable calculational
(usually perturbative, many-body) scheme. This discussion also
alludes to several other interesting conceptual challenges in the
theory of open quantum systems.

Let us presently comment on the issue of the bath-mode
interactions with each other, as well as with impurities, the
latter particularly important and experimentally
relevant\cite{Sakr,JiangPC} for conduction electrons as carriers
of the indirect-exchange interaction. Indeed, the traditional
treatment of open quantum systems has assumed noninteracting bath
modes. When the bath mode interactions had to be accounted for,
\red{the added effects were treated either perturbatively,\cite{MPV}
or, for strong interaction, such as Luttinger-liquid electrons in
a 1D channel, the collective excitations were taken\cite{MDP} as the
new ``bath modes.''}
Generally, however, especially in Markovian
schemes, one has to seek approaches that do not involve certain
double-counting. Indeed, the assumption that the bath modes are at
a fixed temperature, could be possibly considered as partially
accounting for the effects of the mode-mode and mode-impurity
interactions, because these interactions can contribute to
thermalization of the bath, on par with other influences external
to the bath. This may be particularly relevant for phonons that
always have strong anharmonicity for any real material. In a way
this problem fits with the previous one: we are dealing with
effects that can be, on one hand, modeled by added terms in the
total Hamiltonian but on the other hand, may be also mixed in the
process of thermalization that is modeled by actually departing
from the Hamiltonian description and replacing it with Liouville
equations that include noise effects. While all this sounds
somewhat ``foundational,'' recent experimental advances,
interestingly, bring these challenges to the level of application
that requires tractable, albeit perhaps phenomenological solutions
that can be directly confronted with experimental data.

There are other interesting topics to be considered, for instance,
the question of whether additional sources of quantum noise are
possible? \red{For instance, it has been recently established\cite{MM} that potential
difference between two leads (reservoirs, or baths, of electrons)
can be a source of quantum noise with the potential difference
playing the role of the temperature parameter.}

As \red{an example} of a more practical issue to be investigated,
let us mention the possible effect of the sample geometry on
phonon and conduction electron induced relaxation and
interactions. \red{The one-dimensional aspect of the electron gas in a
channel has already been explored}.\cite{MDP} Indeed,
electrons are easy to confine by gate potentials. The situation
for phonons, however, is less clear: can geometrical effects
modify, and particularly reduce, their quantum-noise generation
capacity, or change the induced interactions? Our preliminary
studies \red{reviewed} in this article, seem to indicate strong overall
geometry dependence of the exchange interactions. However, the
situation is not entirely clear, especially for the strength of
the noise effects, and requires a full scale exploration because
recent experiments with double dot nanostructures in Si
membranes\cite{ZhangEtAll} suggest that true nanosize confinement
(due to the sample dimensions) of otherwise long-wavelength modes
(in the transverse sample dimensions) is now possible and will
have dramatic effect on the phonon spectrum and, as a result, on
those physical phenomena that depend on the phonon interactions
with electron spins.

\acknowledgments

We thank D.\ Tolkunov for collaborations and useful
discussions, and acknowledge funding by the NSF under grant
DMR-0121146.

\hphantom{A}



\begin{thebibliography}{99}{\frenchspacing

\bibitem{STPs}  D. Solenov, D. Tolkunov, and V. Privman, Phys. Lett. A {\bf 359}, 81 (2006).

\bibitem{STPb}  D. Solenov, D. Tolkunov, and V. Privman, Phys. Rev. B \textbf{75}, 035134 (2007).

\bibitem{Jiang} M. Xiao, I. Martin, E. Yablonovitch, and H. W. Jiang, Nature \textbf{430}, 435
(2004).

\bibitem{Jiang2} M. R. Sakr, H. W. Jiang, E. Yablonovitch, and E. T. Croke,
Appl. Phys. Lett. \textbf{87}, 223104 (2005).

\bibitem{Craig} N. J. Craig, J. M. Taylor, E. A. Lester, C. M. Marcus, M. P. Hanson, and
A. C. Gossard, Science \textbf{304}, 565 (2004).

\bibitem{Elzerman} J. M. Elzerman, R. Hanson, L. H. Willems van Beveren, B. Witkamp, L. M. K. Vandersypen, and L. P. Kouwenhoven, Nature \textbf{430}, 431 (2004).

\bibitem{Koppens} F. H. L. Koppens, C. Buizert, K. J. Tielrooij, I. T. Vink, K. C.
Nowack, T. Meunier, L. P. Kouwenhoven, and L. M. K. Vandersypen,
Nature {\bf 442}, 766 (2006).

\bibitem{Petta} J. R. Petta, A. C. Johnson, J. M. Taylor, E. A. Laird, A. Yacoby, M. D. Lukin, C. M. Marcus, M. P. Hanson, and A. C. Gossard, Science {\bf 309}, 2180 (2005).

\bibitem{MMJ} I. Martin, D. Mozyrsky, and H. W. Jiang, Phys. Rev. Lett. {\bf 90},
018301 (2003).

\bibitem{PF} E. Prati, M. Fanciulli, A. Kovalev,
J. D. Caldwell, C. R. Bowers, F. Capotondi, G. Biasiol, and L.
Sorba, IEEE Trans. Nanotechnol. {\bf 4}, 100 (2005).

\bibitem{Nakamura} Y. Nakamura, Y. A. Pashkin, and J. S. Tsai, Nature (London) \textbf{398},
786 (1999).

\bibitem{Vion} D. Vion, A. Aassime, A. Cottet, P. Joyez, H. Pothier, C. Urbina,
D. Esteve, and M. H. Devoret, Science \textbf{296}, 886 (2002).

\bibitem{Chiorescu} I. Chiorescu, Y. Nakamura, C. J. P. Harmans, and J. E. Mooij,
Science \textbf{299}, 1869 (2003).

\bibitem{Yamamoto} T. Yamamoto, Y. A. Pashkin, O. Astafiev, Y. Nakamura, and J. S.
Tsai, Nature (London) \textbf{425}, 941 (2003).

\bibitem{Wootters1} S. Hill and W. K. Wootters, Phys. Rev. Lett. \textbf{78}, 5022 (1997).

\bibitem{Wootters2} W. K. Wootters, Phys. Rev. Lett. \textbf{80}, 2245 (1998).

\bibitem{Mahan} G. D. Mahan, \emph{Many-Particle Physics\/} (Kluwer Academic, New York, 2000).

\bibitem{Hasegawa} H. Hasegawa, Phys. Rev. \textbf{118}, 1523 (1960).

\bibitem{Roth} L. M. Roth, Phys. Rev. \textbf{118}, 1534 (1960).

\bibitem{SO-Winkler} R. Winkler, \emph{Spin-Orbit Coupling Effects in Two-Dimentional
Electron and Hole Systems\/} (Springer, New York, 2003).

\bibitem{PGCZ} T. Pellizzari, S. A. Gardiner, J. I. Cirac, and P.
Zoller, Phys. Rev. Lett. \textbf{75}, 3788 (1995).

\bibitem{MKGB} D. Mozyrsky, Sh. Kogan, V. N. Gorshkov, and G. P. Berman, Phys. Rev. B \textbf{65}, 245213 (2002).

\bibitem{PBS} M. Asheghi, Y. K. Leung, S. S. Wong, and K. E. Goodson, Appl. Phys. Lett. \textbf{71}, 1798 (1997).

\bibitem{Leibfried} D. Leibfried, R. Blatt, C. Monroe, and D. Wineland, Rev. Mod. Phys. \textbf{75}, 281 (2003).

\bibitem{Marquet} C. Marquet, F. Schmidt-Kaler, and D. F. V. James, Appl. Phys. B {\bf 76}, 199 (2003).

\bibitem{Porras} D. Porras and J. I. Cirac, Phys. Rev. Lett. \textbf{92}, 207901 (2004).

\bibitem{Leggett} A. J. Leggett, S. Chakravarty, A. T. Dorsey, M. P. A. Fisher, A.
Garg, and W. Zwerger, Rev. Mod. Phys. \textbf{59}, 1 (1987).

\bibitem{Privman}  V. Privman, Modern Phys. Lett. B \textbf{16}, 459 (2002).

\bibitem{Tolkunov2}  D. Tolkunov and V. Privman, Phys. Rev. A \textbf{69}, 062309 (2004).

\bibitem{Solenov}  D. Solenov and V. Privman, Int. J. Modern Phys. B \textbf{20}, 1476 (2006).

\bibitem{Louisell} W. H. Louisell, \emph{Quantum Statistical Properties of Radiation\/} (Wiley, New York, 1973).

\bibitem{PALMA} G. M. Palma, K.-A. Suominen, and A. K. Ekert, Proc. R. Soc. London Ser. A \textbf{452}, 576 (1996).

\bibitem{VKampen} N. G. van Kampen, \emph{Stochastic Processes in Physics and Chemistry\/} (North-Holland, Amsterdam, 2001).

\bibitem{Hanggi} R. Doll, M. Wubs, P. H\"{a}nggi, and S. Kohler, e-print: cond-mat/0703075
(at www.arxiv.org).

\bibitem{Bennett} C. H. Bennett, D. P. DiVincenzo, J. A. Smolin, and W. K. Wootters, Phys. Rev. A \textbf{54}, 3824 (1996).

\bibitem{Vedral} V. Vedral, M. B. Plenio, M. A. Rippin, and P. L. Knight, Phys. Rev. Lett. \textbf{78}, 2275 (1997).

\bibitem{Eberly1} T. Yu and J. H. Eberly, Phys. Rev. B \textbf{68}, 165322 (2003).

\bibitem{Eberly2} T. Yu and J. H. Eberly, Phys. Rev. Lett. \textbf{93}, 140404 (2004).

\bibitem{Braun} D. Braun, Phys. Rev. Lett. \textbf{89}, 277901 (2002).

\bibitem{Blum} K. Blum, \emph{Density Matrix Theory and Applications\/} (Plenum Press, New York, 1996).

\bibitem{Abragam} A. Abragam, \emph{Principles of Nuclear Magnetism} (Clarendon Press, 1983).

\bibitem{Slichter} C. P. Slichter, \emph{Principles of Magnetic Resonance} (Springer, 1991).

\bibitem{FFPReview} A. Fedorov, L. Fedichkin, and V. Privman, J. Comp. Theor. Nanosci. \textbf{1}, 132 (2004).

\bibitem{Privman2} V. Privman, J. Stat. Phys. \textbf{110}, 957 (2003).

\bibitem{Paladino} E. Paladino, M. Sassetti, G. Falci, and U. Weiss, Chem. Phys. \textbf{322}, 98 (2006).

\bibitem{Sakr} M. R. Sakr, E. Yablonovitch, E. T. Croke, and H. W. Jiang, e-print: cond-mat/0504046
(at www.arxiv.org).

\bibitem{JiangPC} H. W. Jiang, private communication.

\bibitem{MPV} D. Mozyrsky, V. Privman, and I. D. Vagner, Phys. Rev. B \textbf{63}, 085313 (2001).

\bibitem{MDP} D. Mozyrsky, A. Dementsov, and V. Privman, Phys. Rev. B \textbf{72}, 233103 (2005).

\bibitem{MM} D. Mozyrsky and I. Martin, Phys. Rev. Lett. \textbf{89}, 018301 (2002).

\bibitem{ZhangEtAll} P. Zhang, E. Tevaarwerk, B.-N. Park, D. E. Savage, G. K. Celler, I. Knezevic, P. G. Evans, M. A. Eriksson, and M. G. Lagally, Nature \textbf{439}, 703 (2006).



 }\end{thebibliography}
\end{document}